\documentclass[aps,pra,twocolumn,showpacs,superscriptaddress]{revtex4}
\usepackage{times,amsmath,amssymb,amsbsy,epsfig,color,graphicx,multirow}

\begin{document}

\title{Teleportation of squeezing: Optimization using non-Gaussian resources}

\author{Fabio Dell'Anno}
\affiliation{Dipartimento di Matematica e Informatica, Universit\`a
degli Studi di Salerno, CNISM, Unit\`a di Salerno, and INFN,
Sezione di Napoli - Gruppo Collegato di Salerno, Via Ponte don Melillo,
I-84084 Fisciano (SA), Italy}

\author{Silvio De Siena}
\affiliation{Dipartimento di Matematica e Informatica, Universit\`a
degli Studi di Salerno, CNISM, Unit\`a di Salerno, and INFN,
Sezione di Napoli - Gruppo Collegato di Salerno, Via Ponte don Melillo,
I-84084 Fisciano (SA), Italy}

\author{Gerardo Adesso}
\affiliation{School of Mathematical Sciences,
University of Nottingham, University Park, Nottingham NG7 2RD, United Kingdom}

\author{Fabrizio Illuminati}
\thanks{Corresponding author. Electronic address:
illuminati@sa.infn.it}
\affiliation{Dipartimento di Matematica e Informatica, Universit\`a
degli Studi di Salerno, CNISM, Unit\`a di Salerno, and INFN,
Sezione di Napoli - Gruppo Collegato di Salerno, Via Ponte don Melillo,
I-84084 Fisciano (SA), Italy}

\date{November 24, 2010}

\begin{abstract}
We study the continuous-variable quantum teleportation of states, statistical moments of
observables, and scale parameters such as squeezing. We investigate the problem both in ideal and
imperfect Vaidman-Braunstein-Kimble protocol setups. We show how the teleportation fidelity is maximized
and the difference between output and input variances is minimized by using suitably optimized
entangled resources. Specifically, we consider the teleportation of coherent
squeezed states, exploiting squeezed Bell states as entangled resources.
This class of non-Gaussian states, introduced in References \cite{CVTelepNoi,RealCVTelepNoi},
includes photon-added and photon-subtracted squeezed states as special cases.
At variance with the case of entangled Gaussian resources, the use of entangled
non-Gaussian squeezed Bell resources allows one to choose different optimization procedures
that lead to inequivalent results. Performing two independent optimization procedures
one can either maximize the state teleportation fidelity, or minimize
the difference between input and output quadrature variances. The two different procedures
are compared depending on the degrees of displacement and squeezing of the input states
and on the working conditions in ideal and non-ideal setups.
\end{abstract}


\maketitle

\section{Introduction}

Non-Gaussian quantum states, endowed with properly enhanced nonclassical properties, may constitute powerful resources for the efficient implementation of quantum information, communication, computation and metrology tasks
\cite{CVTelepNoi,RealCVTelepNoi,KimBS,DodonovDisplnumb,Cerf,Opatrny,Cochrane,Olivares,KitagawaPhotsub,CVTelepNoisyNoi,YangLi,QEstimNoi,QCMenicucci}. Indeed, it has been shown that, at fixed first and second moments, Gaussian states \emph{minimize} various nonclassical properties \cite{ExtremalGaussian,Genoni}. Therefore, many theoretical and experimental efforts have been made towards engineering and controlling highly nonclassical, non-Gaussian states of the radiation field (for a review on quantum state engineering, see e.g.~\cite{PhysRep}). In particular, several  proposals for the generation of non-Gaussian states have been presented \cite{CxKerrKorolkova,AgarTara,DeGauss1,DeGauss2,DeGauss3,DeGauss4,DeGauss5},
and some successful ground-breaking experimental realizations have been already performed \cite{ZavattaScience,ExpdeGauss1,ExpdeGauss2,Grangier,BelliniProbing,GrangierCats}.
Concerning continuous-variable (CV) quantum teleportation,
to date the experimental demonstration of the Vaidman-Braunstein-Kimble (VBK) teleportation protocol \cite{Vaidman,BraunsteinKimble} has been reported both for input coherent states \cite{FurusawaTelep,BowenTelep,ZhangTelep,NoCloningTelep,TeleportationMatter},
and for squeezed vacuum states \cite{TakeiTelep,YonezawaTelep}.
In particular, Ref.~\cite{YonezawaTelep} has reported the teleportation of squeezing,
and consequently of entanglement, between upper and lower sidebands of the same spatial mode.
It is worth to remark that the efficient teleportation of squeezing, as well as of entanglement,
is a necessary requirement for the realization of a quantum information network
based on multi-step information processing \cite{QICV}.

In this paper, adopting the VBK protocol, we study in full generality, e.g.~including loss mechanisms and non-unity gain regimes,
the teleportation of input single-mode coherent squeezed states using as non-Gaussian entangled resources a class of non-Gaussian entangled quantum states, the class of squeezed Bell states \cite{CVTelepNoi,RealCVTelepNoi}.
This class includes, for specific choices of the parameters, non-Gaussian
photon-added and photon-subtracted squeezed states. In tackling our goal, we use the formalism of the characteristic function introduced in Ref.~\cite{MarianCVTelep} for an ideal protocol, and extended to the non-ideal instance in Ref.~\cite{RealCVTelepNoi}. Here, in analogy with the teleportation of coherent states, we first optimize the teleportation fidelity, that is, we look for the maximization of the overlap between the input and the output states. But the presence of squeezing in the unknown input state to be teleported prompts also an alternative procedure, depending on the physical quantities of interest. In fact, if one cares about reproducing in the most faithful way the initial state in phase-space, then the fidelity is
the natural quantity that needs to be optimized. On the other hand, one can be interested in preserving as much as possible the squeezing degree at the output of the teleportation process, even at the expense of
the condition of maximum similarity between input and output states. In this case, one aims at minimizing the difference between the output and input quadrature averages and the quadrature variances. It is important to observe that this distinction makes sense only if one exploits non-Gaussian entangled resources endowed with tunable free parameters, so that enough flexibility is allowed to realize different optimization schemes. Indeed, it is straightforward to verify that this is impossible using Gaussian entangled resources.
We will thus show that exploiting non-Gaussian resources one can identify the best strategies for the
optimization of different tasks in quantum teleportation, such as state teleportation vs teleportation of squeezing. Comparison with the same protocols realized using Gaussian resources will confirm the greater effectiveness of non-Gaussian states vs Gaussian ones as entangled resources in the teleportation
of quantum states of continuous variable systems.

The paper is organized as follows.
In Section \ref{secQTelep}, we introduce the single-mode input states and the two-mode entangled resources, and we recall the basics of both the ideal and the imperfect VKB quantum teleportation protocols. With respect to the instance of Gaussian resources (twin beam), the further free parameters of the non-Gaussian resource (squeezed Bell state) allow one to undertake an optimization procedure to improve the efficiency of the protocols. In Section \ref{secTelepFidelity} we investigate the optimization procedure based on the maximization of the teleportation fidelity.
We then analyze an alternative optimization procedure leading to the minimization of the difference
between the quadrature variances of the output and input fields. This analysis is carried out in Section \ref{secOptVar}. We show that, unlike Gaussian resources, in the instance of non-Gaussian resources the two procedures lead to different results and, moreover, always allow one to improve on the optimization procedures that can be implemented with Gaussian resources. Finally, in Section \ref{secConcl} we draw our conclusions
and discuss future outlooks.

\section{Teleportation of squeezed coherent states using squeezed Bell resources}
\label{secQTelep}

In this Section, we briefly recall the basics of the ideal and imperfect VBK CV teleportation protocols
(for details see Ref.~\cite{RealCVTelepNoi}). The scheme of the (CV) teleportation protocol is the following.
Alice wishes to send to Bob, who is at a remote location, a quantum state, drawn from a particular set according to a prior probability distribution. The set of input states and the prior distribution are known to Alice and Bob, however the specific state to be teleported that is prepared by Alice remains unknown. Alice and Bob share a resource, e.g.~a two-mode entangled state. The input state and one of the modes of the resource are available for Alice, while the other mode of the resource is sent to Bob. Alice performs a suitable (homodyne) Bell measurement, and communicates the result to Bob exploiting a classical communication channel. Then Bob, depending on the result communicated by Alice, performs a local unitary (displacement) transformation, and retrieves the output teleported state. The non-ideal (realistic) teleportation protocol includes mechanisms of loss and inefficiency: the photon losses occurring in the realistic Bell measurements, and the noise arising in the propagation of optical fields in noisy channels (fibers) when the second mode of the resource is sent to Bob. The photon losses occurring in the realistic Bell measurements are modeled by placing in front of an ideal detector a fictitious beam splitter with non-unity transmissivity $T^{2}$ (and corresponding non-zero reflectivity $R^{2}=1-T^{2}$) \cite{LeonhardtRealHomoMeasur}. The propagation in fiber is modeled by the interaction with a Gaussian bath with an effective photon number $n_{th}$, yielding a damping process with inverse-time rate $\gamma$ \cite{WallsMilburn,DecohReview}. Denoting by $in$ the input field mode, and by $1$ and $2$, respectively, the first and the second mode of the entangled resource, the decoherence due to imperfect photo-detection in the homodyne measurement performed by Alice involves the input field mode $in$, and one mode of the resource, e.g.~mode $1$. Throughout, we assume a pure entangled resource. Indeed, it is simple to verify that considering mixed (impure) resources is equivalent to a consider a suitable nonvanishing detection inefficiency $R$ \cite{RealCVTelepNoi}. The degradation due to propagation in fiber affects the other mode of the resource, e.g.~mode $2$, which has to reach Bob's remote place at the output stage. Denoting now by $\rho_{in} \,=\, |\phi\rangle_{in}\,_{in}\langle \phi|$ and $\rho_{res} \,=\, |\psi\rangle_{12}\,_{12}\langle \psi|$ the projectors corresponding, respectively, to a generic pure input single-mode state and a generic pure two-mode entangled resource, the characteristic function $\chi_{out}$ of the single-mode output field $\rho_{out}$ can be written as \cite{RealCVTelepNoi}:
\begin{equation}
\begin{split}
\chi_{out}(\alpha)&={\rm Tr}[D_{out}(\alpha)\rho_{out}]
\\&=e^{- \Gamma_{\tau,R}|\alpha|^{2}}
\chi_{in}\left(g T \, \alpha \right)
\chi_{res}\left(g T \, \alpha^{*};e^{-\frac{\tau}{2}} \, \alpha\right) ,
\end{split}
\label{chioutfinale}
\end{equation}
where $D_{out}(\alpha) = e^{\alpha a_{out}^{\dag} + \alpha^{*} a_{out}}$ is the Glauber displacement operator,
$\chi_{in}(\alpha) = {\rm Tr}[D_{in}(\alpha)\rho_{in}]$ is the characteristic function of the input state,
$\chi_{res}(\alpha_1,\alpha_2) = {\rm Tr}[D_1(\alpha_1)D_2(\alpha_2)\rho_{res}]$ is the characteristic function of the resource,
$g$ is the gain factor of the protocol \cite{TelepGainBowen},
$\tau \equiv \gamma t$ is the scaled dimensionless time proportional to the fiber propagation length, and the function $\Gamma_{\tau,R}$ is defined as:
\begin{equation}
\Gamma_{\tau,R} \,=\, (1-e^{-\tau})\left(\frac{1}{2}+n_{th}\right)+g^{2}R^{2} \,.
\label{Gammadef}
\end{equation}
We assume in principle to have some knowledge about the characteristics of the experimental apparatus: the inefficiency $R$ (or $T$) of the photo-detectors, and the loss parameters $\tau$ and $n_{th}$ of the noisy communication channel.

We consider as input state a single-mode coherent and squeezed (CS) state $|\psi_{CS}\rangle_{in}$ with unknown squeezing parameter $(\varepsilon=s\; e^{i \varphi})$ and unknown coherent amplitude $\beta$. We then consider as non-Gaussian entangled resource the two-mode squeezed Bell (SB) state $|\psi_{SB}\rangle_{12}$, defined as \cite{CVTelepNoi,RealCVTelepNoi}:
\begin{eqnarray}
|\psi_{CS}\rangle_{in}\!\! &=&\!\! D_{in}(\beta) S_{in}(\varepsilon) |0\rangle_{in} \, ,
\label{CohSqueezSt} \\
&& \nonumber \\
|\psi_{SB}\rangle_{12}\!\! &=&\!\! S_{12}(\zeta) \{\cos\delta |0,0 \rangle_{12} + e^{i \theta} \sin\delta |1,1 \rangle_{12} \} \, .
\label{SqueezBell}
\end{eqnarray}
Here $D_{in}(\beta) = e^{\beta a_{in}^{\dag} + \beta^{*} a_{in}}$ is, as before, the displacement operator, $S_{in}(\varepsilon)=e^{-\frac{1}{2}\varepsilon a_{in}^{\dag 2} + \frac{1}{2} \varepsilon^{*} a_{in}^{2}}$ is the single-mode squeezing operator, $S_{12}(\zeta) = e^{ -\zeta a_{1}^{\dag}a_{2}^{\dag} + \zeta^{*} a_{1}a_{2}}$ is the two-mode squeezing operator $(\zeta=r e^{i\phi})$, with $a_{j}$ denoting the annihilation operator for mode $j$ $(j=in,1,2)$, $|m \, , n \rangle_{12}  \equiv |m \rangle_{1} \otimes |n \rangle_{2}$ is the two-mode Fock state (of modes 1 and 2) with $m$ photons in the first mode and $n$ photons in the second mode, and $\theta$ and $\delta$ are two intrinsic free parameters of the resource entangled state, in addition to $r$ and $\phi$, which can be exploited for optimization. Note that particular choices of the angle $\delta$ in the class of squeezed Bell states Eq.~(\ref{SqueezBell}) allow one to recover different instances of two-mode Gaussian and non-Gaussian entangled states:
for $\delta=0$ the Gaussian twin beam (TwB);
for $\delta = \arccos \left[(\cosh 2r)^{-1/2} \sinh r \right]$ and $\theta=\phi-\pi$
the two-mode photon-added squeezed (PAS) state $|\psi_{PAS}\rangle_{12}$;
for $\delta = \arccos \left[(\cosh 2r)^{-1/2} \cosh r \right]$ and $\theta=\phi-\pi$
the two-mode photon-subtracted squeezed (PSS) state $|\psi_{PSS}\rangle_{12}$.
The last two non-Gaussian states are defined as:
\begin{eqnarray}
&&|\psi_{PAS}\rangle_{12} = (\cosh 2r)^{-1/2} a_{1}^{\dag} a_{2}^{\dag} S_{12}(\zeta) |0,0 \rangle_{12} \, ,
\label{PhotAddSqueez} \\
&& \nonumber \\
&&|\psi_{PSS}\rangle_{12} = (\cosh 2r)^{-1/2} a_{1} a_{2} S_{12}(\zeta) |0,0 \rangle_{12} \, ,
\label{PhotSubSqueez}
\end{eqnarray}
and are already experimentally realizable with current technology \cite{ZavattaScience,ExpdeGauss1,ExpdeGauss2,Grangier,BelliniProbing}.

In the following Section we study, in comparison with the instance of two-mode Gaussian entangled resources, the performance of the optimized two-mode squeezed Bell states when used as entangled resources for the teleportation of input single-mode coherent squeezed states. For completeness, in the same context we make also a comparison with the performance, as entangled resources, of the more specific realizations (\ref{PhotAddSqueez}), (\ref{PhotSubSqueez}).
The characteristic functions of states (\ref{CohSqueezSt}), (\ref{SqueezBell}), (\ref{PhotAddSqueez}),
and (\ref{PhotSubSqueez}) are computed and their explicit expressions are given in Appendix \ref{appendixStates}.

\begin{table}
\begin{tabular}{ccc} \hline  \hline
object & parameter &\ \ \ \ \ \ \ \ \ \ \  description\ \ \ \ \ \ \ \ \ \ \ \  \\ \hline \hline
\multirow{3}{*}{input state} & $\beta$ & displacement \\
& $s$ & squeezing degree \\
& $\varphi$ & squeezing phase \\ \hline \hline
\multirow{4}{*}{two-mode resource state} &  $r$ & squeezing degree \\
& $\phi$ & squeezing phase \\
& $\delta$ & mixing angle \\
& $\theta$ & mixing phase \\ \hline \hline
\multirow{4}{*}{teleportation apparatus} & $\small{R\,{=}\sqrt{1-T^2}}$ &  detection inefficiency \\
& $\tau$ & fiber loss factor\\
& $n_{th}$ & fiber bath temperature\\
& $g$ & gain of the protocol\\ \hline  \hline
\end{tabular}
\caption{Summary of the notation employed throughout this work to describe the different parameters that characterize the input coherent squeezed (CS) states [Eq.~(\ref{CohSqueezSt})], the shared entangled two-mode squeezed Bell (SB) resources [Eq.~(\ref{SqueezBell})], and the characteristics of non-ideal teleportation setups \cite{RealCVTelepNoi}. See text for further details on the role of each parameter.}
\label{tableparam}
\end{table}

For ease of reference, table~\ref{tableparam} provides a summary of the parameters associated with the input states, the shared resources, and the sources of noise in the teleportation protocol.

\section{Optimal teleportation fidelity}
\label{secTelepFidelity}

The commonly used measure to quantify the performance of a quantum teleportation protocol is the fidelity of teleportation \cite{braunsteinopt},
$\mathcal{F} \, = \,{\rm Tr}[\rho_{in}\rho_{out}]$, which amounts to the overlap between a pure input state $\rho_{in}$ and the (generally mixed) teleported state $\rho_{out}$.
In the formalism of the characteristic function the fidelity reads
\begin{equation}
\mathcal{F} =  \frac{1}{\pi} \int d^{2}\alpha \;
\chi_{in}(\alpha) \chi_{out}(-\alpha) \, ,
\label{Fidelitychi}
\end{equation}
where $\chi_{in}(\alpha)$ is the characteristic function of the single-mode input state $\rho_{in}=|\psi_{CS}\rangle_{in}\,_{in}\langle\psi_{CS}|$, Eq.~(\ref{CohSqueezSt}), and $\chi_{out}(\alpha)$ is the characteristic function for the output teleported state, Eq.~(\ref{chioutfinale}). In this Section, we will make use of Eq.~(\ref{Fidelitychi}) to analyze the efficiency of the CV teleportation protocol.

In the instance of non-Gaussian squeezed Bell resources (\ref{SqueezBell}), at fixed squeezing parameter,
the optimization procedure amounts to the maximization of the teleportation fidelity (\ref{Fidelitychi})
over the free parameters of the entangled resource.
It can be shown that the optimal choice for the phases $\phi$ and $\theta$ is $\phi=\pi$ and $\theta=0$.
The analytic expression for the fidelity $\mathcal{F}_{CS}$ of the non-ideal quantum teleportation of coherent squeezed states using squeezed Bell resources reads
\begin{widetext}
\begin{eqnarray}
\mathcal{F}_{CS} = &&\frac{4 }{\sqrt{\Lambda_{1}\Lambda_{2}}}
e^{\frac{\omega_{1}^{2}}{\Lambda_{1}}-\frac{\omega_{2}^{2}}{\Lambda_{2}}}
\left\{ 1+e^{-(2r+\tau)}\sin\delta (\Delta_{2}\cos\delta-\Delta_{1}\sin\delta)   \left[\frac{1}{\Lambda_{1}}\left(1+\frac{2\omega_{1}^{2}}{\Lambda_{1}}\right)
+\frac{1}{\Lambda_{2}}\left(1-\frac{2\omega_{2}^{2}}{\Lambda_{2}}\right)\right] \right. \nonumber \\
&&+\frac{1}{4}e^{-2(2r+\tau)} \Delta_{2}^{2}\sin^{2}\delta
\left[\frac{1}{\Lambda_{1}^{2}}\left(3+\frac{12 \omega_{1}^{2}}{\Lambda_{1}}+\frac{4\omega_{1}^{4}}{\Lambda_{1}^{2}}\right)+
\frac{1}{\Lambda_{2}^{2}}\left(3-\frac{12 \omega_{2}^{2}}{\Lambda_{2}}+\frac{4\omega_{2}^{4}}{\Lambda_{2}^{2}}\right)\right.
 \\
&& \left. \left.
+\frac{2}{\Lambda_{1}\Lambda_{2}}\left(1+\frac{2\omega_{1}^{2}}{\Lambda_{1}}-\frac{2\omega_{2}^{2}}{\Lambda_{2}}
-\frac{4\omega_{1}^{2}\omega_{2}^{2}}{\Lambda_{1}\Lambda_{2}}\right)\right]
\right\} \, , \nonumber
\label{TelepFidelityCohSq}
\end{eqnarray}
\end{widetext}
where, introducing $\tilde{g} = g T$, the quantities $\Lambda_{1}$, $\Lambda_{2}$, $\Delta_{1}$, $\Delta_{2}$, $\omega_{1}$, and $\omega_{2}$ are defined by the following relations:
\begin{eqnarray}
&&\Delta_{1} \,=\, 1+e^{4r}+2e^{\frac{\tau}{2}}(1-e^{4r})\tilde{g}+e^{\tau}(1+e^{4r})\tilde{g}^{2} \,, \nonumber \\
&&\Delta_{2} \,=\, 1-e^{4r}+2e^{\frac{\tau}{2}}(1+e^{4r})\tilde{g}+e^{\tau}(1-e^{4r})\tilde{g}^{2} \,, \nonumber \\
&&\Lambda_{1} \,=\, e^{-2r-\tau}\Delta_{1}+2e^{2s}(1+\tilde{g}^{2})+4\Gamma_{\tau,R} \,, \nonumber \\
&&\Lambda_{2} \,=\, e^{-2r-\tau}\Delta_{1}+2e^{-2s}(1+\tilde{g}^{2})+4\Gamma_{\tau,R} \,, \nonumber \\
&&\omega_{1}^{2} \,=\, (1-\tilde{g})^{2}(\beta-\beta^{*})^{2} \,, \quad
\omega_{2}^{2} \,=\, (1-\tilde{g})^{2}(\beta+\beta^{*})^{2} \,, \nonumber \\
&&\tilde{g} \,=\, g T \,.
\label{relations}
\end{eqnarray}

For different choices of $\delta$ in Eq.~(\ref{TelepFidelityCohSq}), see Section \ref{secQTelep},
one obtains the teleportation fidelities associated to photon-added and photon-subtracted squeezed
resource states.
Let us observe that the fidelity in Eq.~(\ref{TelepFidelityCohSq}) depends both on the input coherent amplitude $\beta$, and on the input single-mode squeezing parameter $s$, while it is independent of the input squeezing phase $\varphi$.
Once again, it is worth stressing that, in the teleportation paradigm, the input state is unknown and
only partial (probabilistic) knowledge on the alphabet of input states is admitted.
It is thus required, in principle, to assume teleportation protocols independent of the input parameters,
as it turns out to be the case for the VBK protocol with Gaussian entangled resources and input coherent states. However, in more general cases, one can study the behavior of the so-called one-shot fidelity,
that is the teleportation fidelity at specific values of the input parameters. Suitable averages of the one-shot fidelity over the set of input states and parameters, according to an assigned prior distribution, will then result in the average quantum teleportation fidelity. The latter quantity can then be confronted with so-called classical fidelity thresholds (benchmarks) that correspond to the maximum achievable average fidelity between the input state (measured by Alice in order to achieve an optimal estimation of it) and the output state (prepared by Bob according to Alice's measurement outcomes), without the use of any shared entanglement \cite{braunsteinopt}. While teleportation benchmarks are available for the cases of coherent input states (with completely unknown $\beta$) \cite{BenchmarkCoherent}, purely squeezed input states (with $\beta = \varphi=0$ and completely unknown $s$) \cite{BenchmarkSqueezedNoi}, as well as for states with known squeezing degree and unknown displacement and phase \cite{BenchmarkOwari}, a  benchmark for the case of input states with totally unknown displacement and squeezing has not yet been derived, and stands as a challenging problem in quantum estimation theory.

Henceforth, assuming {\it a priori} that the input parameters (displacement and squeezing degree)
are completely random,
we  adopt then  the following approach to optimize the quantum teleportation fidelity.
We exploit a non-unity gain strategy to remove at least the $\beta$-dependence in the one-shot fidelity;
then, we study the behavior of the $\beta$-independent one-shot fidelity for specific values of the input squeezing parameter $s$, in order to identify an effective, $s$-independent approximation.
Indeed, fixing the gain $g$ at the value $g=1/T$ $(\tilde{g}=1)$ in Eq.~(\ref{TelepFidelityCohSq})
yields the $\beta$-independent fidelity $\mathcal{F}_{S}$:
\begin{eqnarray}
\mathcal{F}_{S} \!\!&=&\!\! \frac{4 }{\sqrt{\Lambda_{1}\Lambda_{2}}}
\bigg\{
\frac{e^{-2(2r+\tau)}}{4}\Delta_{2}^{2}\sin^{2}\delta
\left(\frac{3}{\Lambda_{1}^{2}}+
\frac{3}{\Lambda_{2}^{2}}
+\frac{2}{\Lambda_{1}\Lambda_{2}}\right)     \nonumber \\
&&
 +e^{-(2r+\tau)}\sin\delta (\Delta_{2}\cos\delta-\Delta_{1}\sin\delta)
\left(\frac{1}{\Lambda_{1}}
+\frac{1}{\Lambda_{2}}\right) \nonumber \\
&&  +1
 \bigg\} \,,
\label{TelepFidelitySqVac}
\end{eqnarray}
where the quantities $\Lambda_{1}$, $\Lambda_{2}$, $\Delta_{1}$, $\Delta_{2}$, $\omega_{1}$, and $\omega_{2}$ are defined in Eq.~(\ref{relations}).
For different choices of $\delta$ (see Section \ref{secQTelep}),
one obtains the teleportation fidelities associated to the use of different Gaussian and
non-Gaussian entangled resources: the twin beam, the photon-added, and the photon-subtracted squeezed states. For such resources no optimization procedure is possible as $\delta$ is a specific function of $r$.
Instead, the optimization of the fidelity (\ref{TelepFidelitySqVac}) with respect to the free non-Gaussian parameter $\delta$ identifies the optimal squeezed Bell resource associated to the optimal value:
\begin{equation}
\delta_{opt}\!=\!\frac{1}{2}  \begin{array}{c}\!\arctan\!
\left[\!\frac{4\Delta_{2} \Lambda_{1}\Lambda_{2} (\Lambda_{1}+\Lambda_{2})}{4\Delta_{1}\Lambda_{1}\Lambda_{2} (\Lambda_{1}+\Lambda_{2})-e^{-2r-\tau}\Delta_{2}^{2}(3 \Lambda_{1}^{2}+2\Lambda_{1}\Lambda_{2}+3 \Lambda_{2}^{2})}\!\right]\!\end{array}\!\!.
\label{deltaoptfid}
\end{equation}
Let us notice that, for $\tau=n_{th}=R=0$ (ideal protocol) and $s=0$ (input coherent states), Eq.~(\ref{deltaoptfid}) reduces to \cite{CVTelepNoi}:
\begin{equation}
\delta_{opt} = \frac{1}{2}\arctan
\left[ 1+e^{-2r} \right] \,.
\label{deltaoptfid2}
\end{equation}
The displacement-independent one-shot fidelity $\mathcal{F}_{S}$ and the optimal angle $\delta_{opt}$ are still dependent on $s$, the input squeezing.
Unfortunately, the optimization of the non-Gaussian resource based on the choice (\ref{deltaoptfid})
as optimal angle would be practically unfeasible because the input squeezing is not known.
In order to circumvent this problem, we introduce a sub-optimal angle $\delta_{subopt}$ such that
\begin{equation}\label{deltasubopt}
\delta_{subopt} \equiv \delta_{opt}\big|_{s=\bar{s}}\, ,
\end{equation} where $\bar{s}$ is a fixed effective value of the input squeezing chosen,
according to a suitable criterion that will be clarified below, in the range of possible
values of the squeezing parameter $s$.
\begin{figure}[t]
\centering
\includegraphics[width=8.5cm]{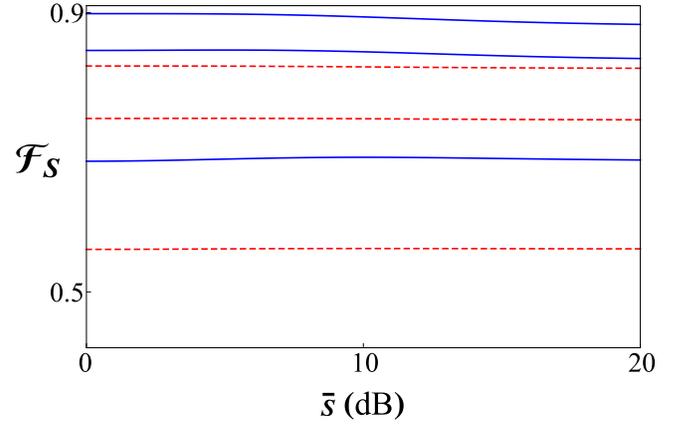}
\caption{(Color online) One-shot fidelity $\mathcal{F}_{S}$ at fixed $s$, and as a function of the angle $\delta$ parameterized by $\bar{s}$, expressed in dB, i.e.~$\delta(\bar{s})=\delta_{subopt}$, see Eq.~(\ref{deltasubopt}), both in the instance of the ideal protocol $\tau = n_{th} = R = 0$ (full lines), and of a non-ideal protocol, with $\tau=0.1$, $n_{th}=0$, and $R^{2}=0.05$ (dashed lines).
The one-shot fidelities are drawn for three different values of the input squeezing: $s=0,\,5,\,10$ dB. The curves are ordered from top to bottom for increasing $s$.}
\label{Fig1sFidsbar}
\end{figure}

In the following we will express the squeezing parameters $r$ and $s$ in decibels,
according to the relation \cite{KokLovett}:
\begin{equation}
\kappa \, (dB) = 10 \log_{10} e^{2\kappa} \,, \qquad \kappa = s,r \,.
\label{sqdecibels}
\end{equation}
The practical rationale for introducing a sub-optimal characterization in the maximization of the output fidelity is based on the observation that the assumption of a completely random degree of input squeezing $s$ is clearly unrealistic. It is instead very sensible to consider that the range of possible values of $s$ falls
in a window $[0,s_{\max}]$ dB. Indeed, to date, the experimentally reachable values of squeezing fall roughly in such a range with $s_{\max} \simeq 10$ dB \cite{MaxSqueezing}.
\begin{figure*}[t]
\centering
\includegraphics[width=17cm]{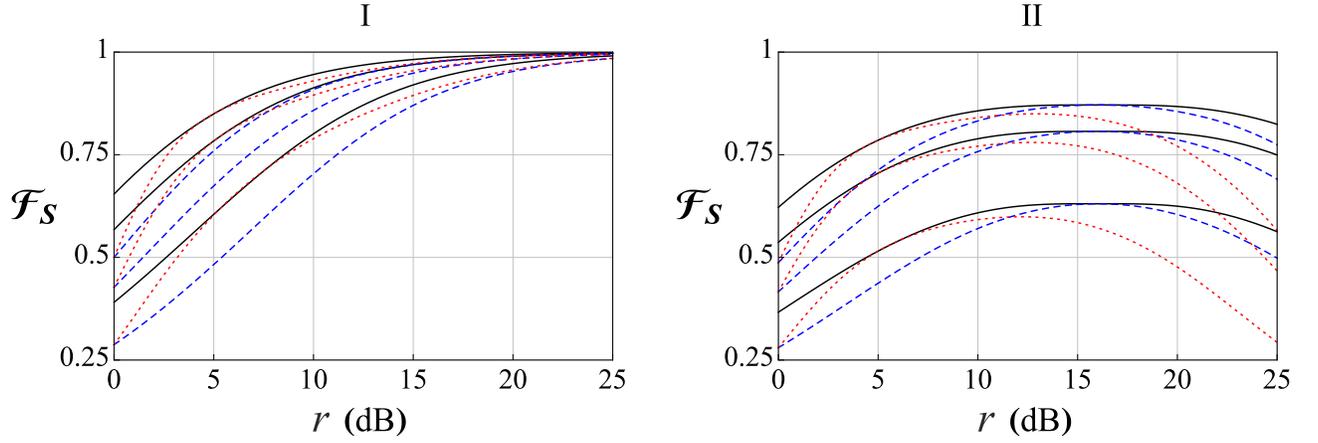}
\caption{(Color online) One-shot fidelity $\mathcal{F}_{S}$ as a function of the squeezing parameter $r$
of the entangled resource, expressed in dB, for the sub-optimal squeezed Bell resource (full black line), for the photon-subtracted squeezed resource (dotted red line), and for the twin beam resource (dashed blue line), in the instance of the ideal protocol (panel I), $\tau = n_{th} = R = 0$, and the non-ideal protocol (panel II), with $\tau=0.1$, $n_{th}=0$, and $R^{2}=0.05$. The one-shot fidelities are drawn for three different values of the input squeezing: $s=0,\,5,\,10$ dB.
In the plots, the fidelity corresponding to the photon-added squeezed resource has been omitted
as it is always lower than the ones corresponding to the photon-subtracted squeezed resource and the squeezed Bell resource. The curves are ordered from top to bottom for increasing $s$.}
\label{Fig1sFid}
\end{figure*}

We can then study the behavior of $\mathcal{F}_{S}$ corresponding to the angle $\delta_{subopt}$ as a function of the effective input squeezing parameter $\bar{s}$, at fixed squeezing parameters of the resource and of the input state, respectively $r$ and $s$, and at fixed loss parameters $\tau$, $n_{th}$, and $R$. Fig.~\ref{Fig1sFidsbar} shows that $\mathcal{F}_{S}$ is quite insensitive to the value of $\bar{s}$. Assuming the realistic range $s \in [0, 10]$ dB, the choice of a sub-optimal angle
such that $\bar{s}=5$ dB (average value of the interval), leads to a decrease of the optimized fidelity, compared to the choice of $\delta_{opt}$, of at most $0.3 \%$ in ideal conditions, and even smaller in realistic conditions. In other words, the teleportation fidelity is essentially constant in the considered interval of variability for the angle $\delta$. Therefore, throughout in the following, we fix $s=\bar{s}=5$ dB in the expression Eq.~(\ref{deltaoptfid}) to make it $s$-independent.

In Fig.~\ref{Fig1sFid}, we plot the teleportation fidelity associated to the various considered resources (Gaussian twin beam, optimized two-mode squeezed Bell-like state, two-mode squeezed photon-subtracted state)
both for the ideal protocol (panel I) and for the non-ideal protocol (panel II).
We see that, at fixed (finite) squeezing $r$ of the resource,
the Gaussian twin beam is always outperformed by the optimal non-Gaussian squeezed Bell resource
in the ideal protocol.
It is worth to remark that for very high values of the squeezing $r$, the advantage of the non-Gaussian resources fades and Gaussian twin beams perform in practice equally well for the teleportation of the considered input states. This reflects the well known fact that, using the ideal VBK protocol and an ideal Einstein--Podolsky--Rosen resource (corresponding, e.g., to a twin beam in the limit $r \rightarrow \infty$), {\it any} quantum state can be unconditionally teleported with unit fidelity \cite{Vaidman}.
All the one-shot fidelities decrease for increasing squeezing $s$ of the input and, interestingly,
in the non-ideal protocol they achieve a maximum at a finite value $r_m$ of the squeezing $r$ of the resource.
The optimal squeezed Bell resource and the twin beam share the same $r_m \simeq 16$ dB and coincide at that point.
In Fig.~\ref{Fig1sFid} we also plot the one-shot fidelities associated with the two-mode
photon-subtracted squeezed states, Eq.~(\ref{PhotSubSqueez}).
The two-mode photon-subtracted squeezed state always outperforms the twin beam in the ideal protocol,
and at low and intermediate values of the resource squeezing $r$ in the non-ideal case. It is always
outperformed by the optimized squeezed Bell resource.
We note that, on the other hand, the two-mode photon-added squeezed states always exhibit a performance worse than the two-mode photon-subtracted squeezed states and the squeezed Bell resources (the corresponding fidelities are omitted in the plots for clarity).
In a given range of the squeezing $r$, $|\psi_{PSS}\rangle_{12}$ and $|\psi_{SB}\rangle_{12}$
exhibit comparable levels in the fidelity of teleportation.
In conclusion, properly optimized non-Gaussian resources maximize the fidelity of teleportation of squeezed coherent states
both in the ideal and imperfect VBK protocols, outperforming the corresponding Gaussian resources.
In the next Section we carry out a similar analysis with the aim of identifying the optimal strategy
that maximizes the reproduction at the output of the input squeezing.

\section{Teleportation of quadrature moments}
\label{secOptVar}

In this Section, we introduce a different approach to the optimization of the teleportation protocol,
aimed at retaining and faithfully reproducing at the output the variances and thus the squeezing of
the input state. The strategy is to constrain the first and second order moments of the output field to reproduce the ones of the input field, by exploiting the free parameters of the non-Gaussian resources.
We introduce the mean values
$\langle Z_j \rangle = {\rm Tr}[Z_j \rho_{j} ]$, with $Z_j = X_j , P_j$ $(j=in,\,out)$,
and the variances
$\langle \Delta Z_{j}^{2} \rangle ={\rm Tr}[ Z_{j}^{2} \rho_{j}] - {\rm Tr}[ Z_j \rho_{j} ]^{2}$, and
$\langle \Delta (X_{j} P_{j})_S \rangle ={\rm Tr}[ (X_{j} P_{j}+P_{j} X_{j}) \rho_{j}] - 2{\rm Tr}[ X_j \rho_{j} ]{\rm Tr}[ P_j \rho_{j} ]$ (the cross-quadrature variance, with $S$ denoting the symmetrization)
of the quadrature operators
$X_{j} \,=\, \frac{1}{\sqrt{2}}(a_{j}+ a_{j}^{\dag})$,
$P_{j} \,=\, \frac{i}{\sqrt{2}}(a_{j}^{\dag}- a_{j})$,
associated with the single-mode input state $\rho_{in}$ and the output state $\rho_{out}$ of the teleportation protocol.
The explicit expressions for the quantities $\langle Z_j \rangle$, $\langle \Delta Z_{j}^{2} \rangle$, and
$\langle \Delta (X_{j} P_{j})_S \rangle$
are reported in the Appendix \ref{appendixQuadratures}.

\begin{figure*}[t]
\centering
\includegraphics[width=18cm]{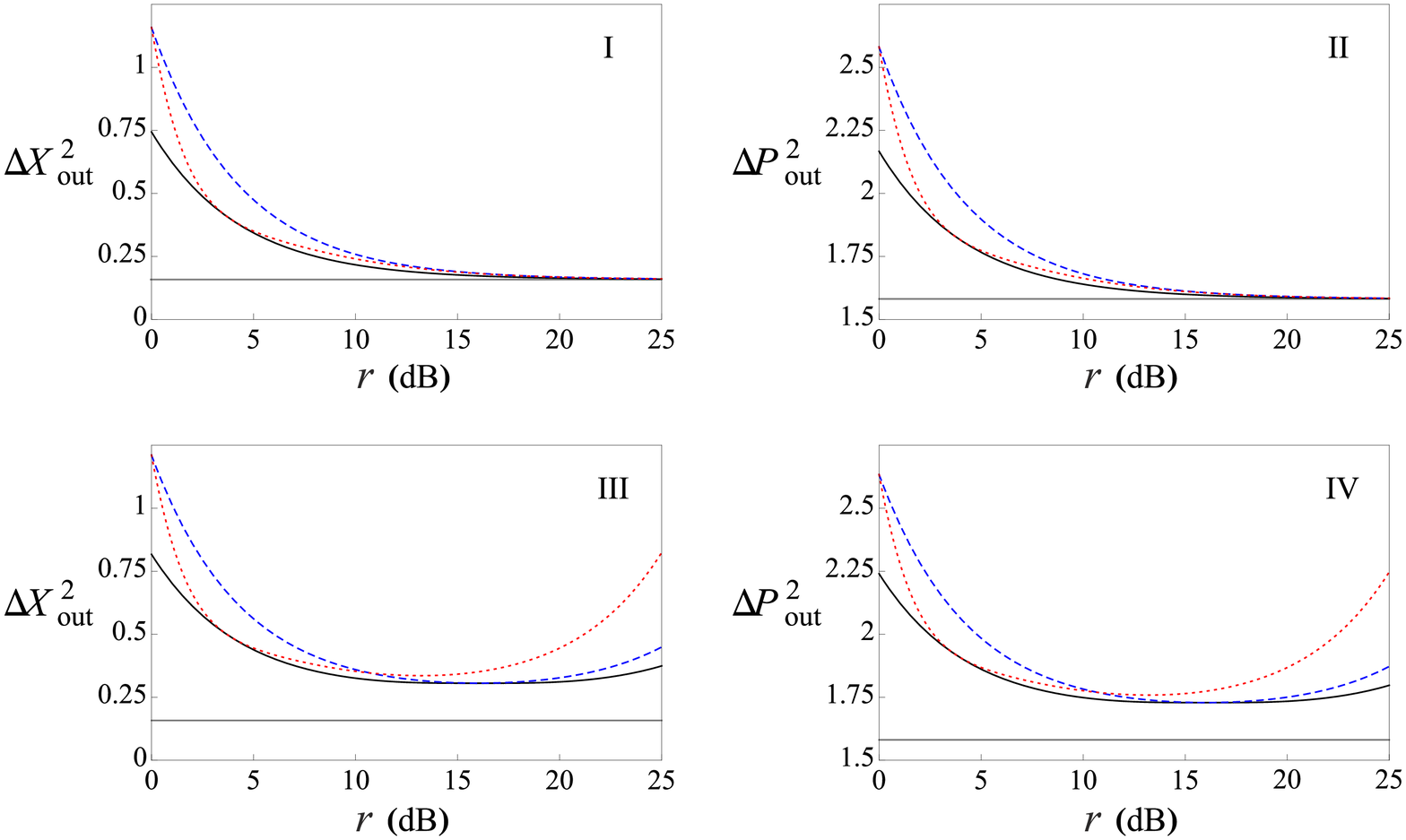}
\caption{(Color online) Output variances $\langle \Delta X_{out}^{2} \rangle$
and $\langle \Delta P_{out}^{2} \rangle$,
as a function of the squeezing parameter $r$ of the resource, expressed in dB,
for the optimized squeezed Bell resource (full black line),
for the photon-subtracted squeezed resource (dotted red line),
and for the twin beam resource (dashed blue line).
Panels I and II: ideal protocol. Panels III and IV:
non-ideal protocol. The various curves are to be compared with
the given input variances of the input single-mode squeezed coherent
state (horizontal solid lines). The squeezing of the input state
is fixed at $s=5$ dB and $\varphi=0$. In the non-ideal protocol,
the experimental parameters are fixed at
$\tau=0.1$, $n_{th}=0$, and $R^{2}=0.05$.
In the plots, the variances associated with the photon-added squeezed resource have been omitted
as they are always larger than the ones corresponding to the photon-subtracted squeezed resource and the squeezed Bell resource.}
\label{FigVar}
\end{figure*}

The quantities measuring the deviation of the output from the input are the
differences between the output and input first and second quadrature moments:
\begin{eqnarray*}
&&D(X) \equiv \langle X_{out} \rangle -\langle X_{in} \rangle = (\tilde{g}-1) \langle X_{in} \rangle \,,
\label{deltameanx} \\
&&D(P) \equiv \langle P_{out} \rangle -\langle P_{in} \rangle = (\tilde{g}-1) \langle P_{in} \rangle \,,
\label{deltameanp} \\
&&D(\Delta X^{2}) \equiv \langle \Delta X_{out}^{2} \rangle - \langle \Delta X_{in}^{2}\rangle =
(\tilde{g}^{2}-1) \langle \Delta X_{in}^{2}\rangle +\Sigma \,,
\label{deltavarx} \\
&&D(\Delta P^{2}) \equiv \langle \Delta P_{out}^{2} \rangle - \langle \Delta P_{in}^{2}\rangle =
(\tilde{g}^{2}-1) \langle \Delta P_{in}^{2}\rangle +\Sigma \,,
\label{deltavarp} \\
&&D(\Delta (X P)_S) \equiv \langle \Delta (X_{out} P_{out})_S \rangle - \langle \Delta (X_{in} P_{in})_S \rangle
\nonumber \\
&& = (\tilde{g}^{2}-1) \langle \Delta (X_{in} P_{in})_S \rangle \,,
\label{deltavarxp}
\end{eqnarray*}
\noindent with $\Sigma$ given by Eq.~(\ref{EqC}).
From the above equations, we see that the assumption $\tilde{g}=1$ (i.e.~$g=1/T$) yields $D(X)=D(P)=D(\Delta (X P)_S) =0$ and
$D(\Delta X^2)=D(\Delta P^2)=\Sigma\big|_{\tilde{g}=1}$.
Therefore, for $\tilde{g}=1$, the input and output fields possess equal average position and momentum (equal first moments),and equal cross-quadrature variance;
then, the optimization procedure reduces to the minimization of the quantity $\Sigma\big|_{\tilde{g}=1}$ with respect to the free parameters of the non-Gaussian squeezed Bell resource, i.e.~$\min_{\delta,\theta,\phi}\Sigma\big|_{\tilde{g}=1}$.
Moreover, as for the optimization procedure of Section \ref{secTelepFidelity},
it can be shown that the optimal choice for $\phi$ and $\theta$ is, once again, $\phi=\pi$ and $\theta=0$.
The optimization on the remaining free parameter $\delta$ yields the optimal value $\delta_{optvar}$:
\begin{equation}
\delta_{optvar} = \frac{1}{2}\arctan \left[\frac{(1+e^{\frac{\tau}{2}})^2-e^{4r}(1-e^{\frac{\tau}{2}})^2}{(1+e^{\frac{\tau}{2}})^2
+e^{4r}(1-e^{\frac{\tau}{2}})^2}\right] \, .
\label{deltaoptvar}
\end{equation}
The optimal angle $\delta_{optvar}$, corresponding to the minimization of the differences $D(\Delta X^2)$ and $D(\Delta P^2)$ between the output and input quadrature variances, is independent of $R$, at variance with the optimal value $\delta_{opt}$, Eq.~(\ref{deltaoptfid}), corresponding to the maximization of the teleportation fidelity. It is also important to note that in this case there are no questions related to a dependence on the input squeezing $s$. For $\tau=0$ Eq.~(\ref{deltaoptvar}) reduces to $\delta_{optvar} = \pi/8$. Such a value is equal to the asymptotic value given by Eq.~(\ref{deltaoptfid2}) for
$r \rightarrow \infty$, so that, in this extreme limit the two optimization procedures become equivalent.
In the particular cases of photon-added and photon-subtracted resources, no optimization procedure can be carried out, and the parameter $\delta$ is simply a given specific function of $r$ (see Section \ref{secQTelep}). We remark that, having automatically zero difference in the cross-quadrature variance at $\tilde{g}=1$, finding the angles that minimize $D(\Delta X^2)$ and $D(\Delta P^2)$ precisely solves the problem of achieving the optimal teleportation of both the first moments and the full covariance matrix of the input state at once.

In order to compare the performances of the Gaussian and non-Gaussian resources,
and to emphasize the improvement of the efficiency of teleportation with squeezed Bell-like states,
we consider first the instance of ideal protocol ($\tau=0$, $n_{th}=0$, $R=0$), and compute,
and explicitly report below, the output variances $\langle \Delta Z_{out}^{2} \rangle_{f}$
of the teleported state associated with non-Gaussian resources
(i.e.~optimized squeezed Bell-like states $(f=SB)$, photon-added squeezed states $(f=PAS)$,
photon-subtracted squeezed states $(f=PSS)$), and with Gaussian resources, i.e.~twin beams $(f=TwB)$.
From Eqs.~(\ref{VarXout})--(\ref{EqC3}), we get:
\begin{eqnarray}
\langle \Delta Z_{out}^{2} \rangle_{TwB}\!&=&\!\langle \Delta Z_{in}^{2} \rangle + e^{-2r} \,.
\label{DZTwB}
\\ && \nonumber \\
\langle \Delta Z_{out}^{2} \rangle_{PAS}\!&=&\!\langle \Delta Z_{in}^{2} \rangle + e^{-2r} \left\{1+\frac{2e^{-2r}(1+e^{-2r})}{1+e^{-4r}}\right\} \!,\nonumber
\\ &&\label{DZPAS}
\\ && \nonumber \\
\langle \Delta Z_{out}^{2} \rangle_{PSS}\!&=&\!\langle \Delta Z_{in}^{2} \rangle + e^{-2r} \left\{1-\frac{2e^{-2r}(1-e^{-2r})}{1+e^{-4r}}\right\} \!,\nonumber
\\
 && \label{DZPSS}
\\ && \nonumber \\
\langle \Delta Z_{out}^{2} \rangle_{SB}\!&=&\!\langle \Delta Z_{in}^{2} \rangle + e^{-2r} (2-\sqrt{2}) \,,
\label{DZSB}
\end{eqnarray}
Eq.~(\ref{DZSB}) is derived exploiting the optimal angle (\ref{deltaoptvar}),
which reduces to Eq.~(\ref{deltaoptfid2}) in the ideal case.
Independently of the resource, the teleportation process will in general result
in an amplification of the input variance. However, the use of non-Gaussian optimized
resources, compared to the Gaussian ones, reduces sensibly the amplification
of the variances at the output. Looking at Eq.~(\ref{DZTwB}), we see that the teleportation
with the twin beam resource produces an excess, quantified by the exponential term $e^{-2r}$,
of the output variance with respect to the input one. On the other hand, the use of
the non-Gaussian squeezed Bell resource Eq.~(\ref{DZSB})
yields a reduction in the excess of the output variance with respect to the input one by a factor $(2-\sqrt{2})$.
Let us now analyze the behaviors of the photon-added squeezed resources
and of the photon-subtracted squeezed resources, Eqs.~(\ref{DZPAS}) and (\ref{DZPSS}), respectively.
We observe that, in analogy with the findings of the previous Section,
the photon-subtracted squeezed resources exhibit an intermediate behavior in the ideal protocol;
indeed for low values of $r$ they perform better than the Gaussian twin beam,
but worse than the optimized squeezed Bell states.
The photon-added squeezed resources perform worse than both the twin beam and the other non-Gaussian resources.
These considerations  follow straightforwardly from a quantitative analysis of the terms
associated with the excess of the output variance in Eqs.~(\ref{DZPAS}) and (\ref{DZPSS}).
Moreover, again in analogy with the analysis of the optimal fidelity, for low values of $r$,
there exists a region in which the performance of photon-subtracted squeezed states
and optimized squeezed Bell states are comparable.
Finally, again in analogy with the case of the fidelity optimization, the output variance
associated with the Gaussian twin beam and with the optimized squeezed Bell states coincide
at a specific, large value of $r$, at which the two resources become identical.

The input variances $\langle \Delta Z_{in}^{2}\rangle$ (\ref{VarXin}) and (\ref{VarPin}),
and the output variances $\langle \Delta Z_{out}^{2} \rangle$,
are plotted in panels I and II of Fig.~\ref{FigVar} for the ideal VKB protocol
and in panels III and IV of Fig.~\ref{FigVar} for the non-ideal protocol.

In the instance of realistic protocol, for small resource squeezing degree $r$, similar conclusions can be drawn, leading to the same hierarchy among the entangled resources.
However, analogously to the behavior of the teleportation fidelity, for high values of $r$
the photon-subtracted squeezed resources are very sensitive to decoherence.
In fact, such resources perform worse and worse than the Gaussian twin beam
for $r$ greater than a specific finite threshold value.

Rather than minimizing the differences between output and
input quadrature variances, one might be naively tempted to
consider minimizing the difference between the ratio of the
output variances $\langle \Delta X_{out}^{2} \rangle / \langle \Delta P_{out}^{2} \rangle$
and the ratio of the input variances
$\langle \Delta X_{in}^{2} \rangle / \langle \Delta P_{in}^{2} \rangle$.
This quantity might appear to be of some interest because it is a good measure of
how well squeezing is teleported in all those cases in which the input and output
quadrature variances are very different, that is those situations in which
the statistical moments are teleported with very low efficiency. However, it is
of little use to preserve formally a scale parameter if the noise on the quadrature
averages grows out of control. The procedure of minimizing the difference between
output and input quadrature statistical moments is the only one that guarantees
the simultaneous preservation of the squeezing degree and the reduction of the
excess noise on the output averages and statistical moments of the field observables.

\section{Conclusions}\label{secConcl}

We have studied the efficiency of the VBK CV quantum teleportation protocol for the transmission of quantum  states and averages of observables using optimized non-Gaussian entangled resources.
We have considered the problem of teleporting Gaussian squeezed and coherent states, i.e.~input
states with two unknown parameters, the coherent amplitude and the squeezing.
The non-Gaussian resources (squeezed Bell states) are endowed with free parameters that can be tuned to
maximize the teleportation efficiency either of the state or of physical quantities such as squeezing,
quadrature averages, and statistical moments. We have discussed two different optimization procedures:
the maximization of the teleportation fidelity of the state, and the optimization of the teleportation
of average values and variances of the field quadratures. The first procedure maximizes the similarity in phase space between the teleported and the input state, while the second one maximizes the preservation
at the output of the displacement and squeezing contents of the input.

We have shown that optimized non-Gaussian entangled resources such as the squeezed Bell states,
as well as other more conventional non-Gaussian entangled resources,
such as the two-mode squeezed photon-subtracted states,
outperform, in the realistic intervals of the squeezing parameter $r$ of the entangled resource
achievable with the current technology,
entangled Gaussian resources both for the maximization of the teleportation fidelity and
for the maximal preservation of the input squeezing and statistical moments. These findings are consistent
and go in line with previous results on the improvement of various quantum information protocols replacing
Gaussian with suitably identified non-Gaussian resources \cite{CVTelepNoi,RealCVTelepNoi,Opatrny,Cochrane,Olivares,KitagawaPhotsub}.
In the process, we have found that the two optimal values of the resource angle $\delta$ associated with the two optimization procedures are different and identified, respectively, by
Eqs.~(\ref{deltaoptfid}) and (\ref{deltaoptvar}). This inequivalence is connected to the fact that,
when using entangled non-Gaussian resources with free parameters that are amenable to optimization,
the fidelity is closely related to the form of the different input properties that one wishes to teleport,
e.g.~quasi-probability distribution in the phase space, squeezing, statistical moments of higher order,
and so on. Different quantities correspond to different optimal teleportation strategies.

Finally, regarding the VBK protocol, it is worth remarking that the maximization of the teleportation fidelity corresponds to the maximization of the squared modulus of the overlap
between the input and the output (teleported) state, without taking into account the
characteristics of the output with respect to the input state.
Therefore, part of the non-Gaussian character of the entangled resource is unavoidably
transferred to the output state. The latter then acquires unavoidably a certain degree
of non-Gaussianity, even if the presence of pure Gaussian inputs.
Moreover, as verified in the case of non-ideal protocols, the output state is also strongly
affected by decoherence. Thus, in order to recover the purity and the Gaussianity of the teleported state,purification and Gaussification protocols should be implemented serially after transmission
through the teleportation channel is completed \cite{GaussificSchemes}. If the second (squeezing preserving) procedure is instead considered, the possible deformation of the Gaussian character is not so relevant, because the shape reproduction is not the main goal, while purification procedures are again needed to correct for the extra noise added during teleportation when finite entanglement and realistic conditions are considered.

An important open problem is determining a proper teleportation benchmark for the class of Gaussian input states with unknown displacement and squeezing. Such a benchmark is expected to be certainly smaller than $50\%$ in terms of teleportation fidelity, the latter being the benchmark for purely coherent input states with completely random displacement in phase space \cite{braunsteinopt,BenchmarkCoherent}.  Our results indicate that optimized non-Gaussian entangled resources will allow one to beat the classical benchmark, thus achieving unambiguous quantum state transmission via a truly quantum teleportation, with a smaller amount of nonclassical resources, such as squeezing and entanglement, compared to the case of shared Gaussian twin beam resources. In this context, Fig.~\ref{Fig1sFid} provides strong and encouraging evidence that suitable uses of non-Gaussianity in tailored resources, feasible with current technology \cite{ZavattaScience,ExpdeGauss1,ExpdeGauss2,Grangier,BelliniProbing}, may lead to a genuine  demonstration of CV quantum teleportation of displaced squeezed states in realistic conditions of the experimental apparatus. This would constitute a crucial step forward after the successful recent experimental achievement of the quantum storage of a displaced squeezed thermal state of light into an atomic ensemble memory \cite{polzikalessio}.

\vspace{0.5cm}

\acknowledgments
We acknowledge financial support from the European Union under the FP7 STREP Project HIP
(Hybrid Information Processing), Grant Agreement No. 221889.


\vspace{1cm}

\appendix

\section{Input states, entangled resources, and output states}
\label{appendixStates}

Here we report the characteristic functions for the single-mode input states and for the two-mode entangled resources. The characteristic function for the coherent squeezed states (\ref{CohSqueezSt}), i.e.~$\chi_{in}(\alpha)=\,_{in}\langle\psi_{CS}| D_{in}(\alpha)|\psi_{CS}\rangle_{in}\,,$ reads:
\begin{equation}
\chi_{in}(\alpha)=e^{\frac{1}{2}(\alpha \beta^{*}-\alpha^{*}\beta)-\frac{1}{2}\left|(\alpha+\beta)\cosh s+(\alpha^{*}+\beta^{*})e^{i\varphi}\sinh s\right|^{2}} \,.
\label{chiinput}
\end{equation}
The characteristic function for the squeezed Bell-like resource (\ref{SqueezBell}), i.e.~$\chi_{SB}(\alpha_1 , \alpha_2)=\,_{12}\langle\psi_{SB}| D_1(\alpha_1)D_2(\alpha_2) |\psi_{SB}\rangle_{12}\,,$
reads:
\begin{equation}
\begin{split}
\chi_{SB}(\alpha_{1},\alpha_{2})&=e^{-\frac{1}{2}(|\xi_{1}|^{2}+|\xi_{2}|^{2})} \\
&\times \big[1+\sin\delta\cos\delta(e^{i\theta}\xi_{1}\xi_{2}+e^{-i\theta}\xi_{1}^{*}\xi_{2}^{*}) \\
&\quad +\sin^{2}\delta (|\xi_{1}|^{2}|\xi_{2}|^{2}-|\xi_{1}|^{2}-|\xi_{2}|^{2})\big] \,,
\end{split}
\label{charfuncSB}
\end{equation}
where the complex variables $\xi_{k}$ are defined as:
\begin{equation}
\xi_{k}=\alpha_{k}\cosh r +\alpha_{l}^{*} e^{i\phi} \sinh r \quad (k,l=1,2;k\neq l).
\end{equation}
It is worth noticing that, for $\delta=0$, Eq.~(\ref{charfuncSB}) reduces to the well-known Gaussian characteristic function of the twin beam. Given the characteristic functions for the single-mode the input state and for the two-mode entangled resource, Eqs.~(\ref{chiinput}) and (\ref{charfuncSB}), respectively, it is straightforward to obtain the characteristic function for the single-mode output state of the teleportation protocol by using Eq.~(\ref{chioutfinale}) and replacing $\chi_{res}$ with $\chi_{SB}$.

\section{Mean values and variances of the quadratures}
\label{appendixQuadratures}

In this Appendix, we report the analytical expressions for the mean values $\langle Z_j \rangle = {\rm Tr}[Z_j \rho_{j} ]$,
with $Z_j = X_j , P_j$ $(j=in,\,out)$,
and the variances $\langle \Delta Z_{j}^{2} \rangle ={\rm Tr}[ Z_{j}^{2} \rho_{j}] - {\rm Tr}[ Z_j \rho_{j} ]^{2}$
of the quadrature operators $X_{j} \,=\, \frac{1}{\sqrt{2}}(a_{j}+ a_{j}^{\dag})$,
$P_{j} \,=\, \frac{i}{\sqrt{2}}(a_{j}^{\dag}- a_{j})$, associated with the single-mode input state $\rho_{in}$
and the output state $\rho_{out}$ of the teleportation protocol.
We also compute the cross-quadrature variance
$\langle \Delta (X_{j} P_{j})_S \rangle ={\rm Tr}[ (X_{j} P_{j}+P_{j} X_{j}) \rho_{j}] - 2{\rm Tr}[ X_j \rho_{j} ]{\rm Tr}[ P_j \rho_{j} ]$,
associated with the non-diagonal term of the covariance matrix of the density operator,
where the subscript $S$ denotes the symmetrization.
The mean values and the variances associated with the input single-mode coherent squeezed state (\ref{CohSqueezSt}) can be easily computed:
\begin{eqnarray}
&&\langle X_{in} \rangle = \frac{1}{\sqrt{2}}(\beta+\beta^{*}) , \label{avXin} \\
&&\langle P_{in} \rangle = \frac{i}{\sqrt{2}}(\beta^{*}-\beta), \label{avPin}
\end{eqnarray}
and
\begin{eqnarray}
&&\langle \Delta X_{in}^{2} \rangle = \frac{1}{2} (\cosh 2s - \cos\varphi \sinh 2s) ,
\label{VarXin} \\
&& \nonumber \\
&&\langle \Delta P_{in}^{2} \rangle = \frac{1}{2} (\cosh 2s + \cos\varphi \sinh 2s) ,
\label{VarPin}
\\
&& \nonumber \\
&&\langle \Delta (X_{in} P_{in})_S \rangle = - \sin\varphi \sinh 2s .
\label{VarXinPin}
\end{eqnarray}
The mean values and the variances associated with the output single-mode teleported state,
described by the characteristic function (\ref{chioutfinale}) read:
\begin{eqnarray}
&&\langle X_{out} \rangle = \frac{\tilde{g}}{\sqrt{2}}(\beta+\beta^{*}) , \label{avXout} \\
&&\langle P_{out} \rangle = \frac{i \tilde{g}}{\sqrt{2}}(\beta^{*}-\beta), \label{avPout}
\end{eqnarray}
and
\begin{eqnarray}
&&\langle \Delta X_{out}^{2} \rangle = \tilde{g}^{2} \langle \Delta X_{in}^{2}\rangle  +\Sigma \,,
\label{VarXout} \\
&& \nonumber \\
&&\langle \Delta P_{out}^{2} \rangle = \tilde{g}^{2} \langle \Delta P_{in}^{2}\rangle  +\Sigma \,,
\label{VarPout} \\
&& \nonumber \\
&&\langle \Delta (X_{out} P_{out})_S \rangle = - \tilde{g}^{2} \sin\varphi \sinh 2s \,,
\label{VarXoutPout}
\end{eqnarray}
with
\begin{widetext}
\begin{eqnarray}
\Sigma &=& \Gamma_{\tau,R}+e^{-\frac{\tau}{2}}\tilde{g} \sin(\theta-\phi) \sin\phi \sin 2\delta-\frac{1}{4}e^{-2r-\tau}(1+e^{\tau}\tilde{g}^{2}-2 e^{\frac{\tau}{2}}\tilde{g}\cos\phi)
[\cos 2\delta-\cos(\theta-\phi) \sin 2\delta-2] \nonumber \\
&-&\frac{1}{4}e^{2r-\tau}(1+e^{\tau}\tilde{g}^{2}+2 e^{\frac{\tau}{2}}\tilde{g}\cos\phi)
[\cos 2\delta +\cos(\theta-\phi) \sin 2\delta-2]   \,.
\label{EqC}
\end{eqnarray}
For the particular choices $\tilde{g}=1$, $\phi=\pi$, and $\theta=0$, Eq.~(\ref{EqC}) reduces to:
\begin{equation}
\Sigma\big|_{\tilde{g}=1} = \Gamma_{\tau,R}\big|_{g=1/T}-\frac{1}{4}e^{-2r-\tau}(1+ e^{\frac{\tau}{2}})^{2}
[\cos 2\delta+ \sin 2\delta-2]
-\frac{1}{4}e^{2r-\tau}(1- e^{\frac{\tau}{2}})^{2}
[\cos 2\delta - \sin 2\delta-2]
 \,.
 \label{EqC2}
\end{equation}
\end{widetext}
In the instance of Gaussian resource $(\delta=0)$, such quantity simplifies to:
\begin{equation}
\Sigma_G = \Gamma_{\tau,R}\big|_{g=1/T}+ \frac{1}{4}e^{-2r-\tau}(1+ e^{\frac{\tau}{2}})^{2}
+\frac{1}{4}e^{2r-\tau}(1- e^{\frac{\tau}{2}})^{2}
 \,.
\label{EqC3}
\end{equation}
For suitable choices of $\delta$ in Eq.~(\ref{EqC2}), see Section \ref{secQTelep},
one can easily obtain the output variances associated with photon-added and photon-subtracted squeezed states.

\end{document}